# Generalized Bloch-Wangsness-Redfield Kinetic Equations.

N.F. Fatkullin

Kazan Federal University (former Kazan State University), Institute of Physics
Kremlevskaya 18, 400008 , Kazan, Russia/Tatarstan, e-mail: Nail.Fatkullin@ksu.ru

**Abstract**

We present a compact and general derivation of the generalized Bloch-Wangsness-Redfield kinetic equations for systems with the static spin Hamiltonian utilizing the concept of the Liouville space. We show that the assumptions of short correlation times and large heat capacity of the lattice are sufficient to derive the kinetic equations without the use of perturbation theory for the spin-lattice interaction operator. The perturbation theory is only applied for calculation of the kinetic coefficients, for which we obtain general and compact expressions. We argue that kinetic equations for the density matrix elements are not essential for derivation of the generalized Bloch-Wangsness-Redfield equations for the expectation values of any set of physical quantities, and the latter may be obtained directly under the weak assumptions of mutual orthogonality and completeness. We show that existence of a unity operator in the algebra of spin operators is a necessary condition for convergence of the spin subsystem to thermodynamic equilibrium with the temperature equal to that of the lattice. Finally, as an application of the general theory, we discuss some features of spin relaxation in heterogeneous systems for $I=1/2$.

Keywords: Bloch-Wangsness-Redfield kinetic equations, spin-relaxation.

## 1. Introduction.

Derivation of kinetic equations for spin subsystem that interacts with a "lattice" in thermodynamically equilibrium state is a major problem in the theory of magnetic resonance in particular and is of considerable interest for the general theory of nonequilibrium processes.

For the first time this problem was considered by Bloch, Wangsness, Redfield [1-4]. Subsequent development and analysis of the theory called afterwards as a Bloch – Wangsness – Redfield theory (BWR) are reviewed in a number of works (see for example [5-12]).

The best results are obtained when one uses an approximation of short correlation times for a case of free relaxation of spin subsystem when it interacts with a lattice having large heart capacity. One should feel incompleteness even in this case.

In all known works the kinetic equations have been derived starting from the perturbation theory for the spin-lattice interaction operator. More over the kinetic equations for the elements of nonequilibrium spin density matrix play a central role. The kinetic coefficients obtained appear to be bulky and explicitly "noninvariant".

Further basing on this kinetic equation one must perform additional transformation in order to obtain the generalized Bloch equations for the expectation values of physical quantities.

A proof of thermodynamic equilibrium establishing in spin subsystem with the temperature equal to the lattice temperature seems for us to be uncompleted. In general case the proof comes to nothing than to demonstrate that the equilibrium density matrix is a stationary solution of corresponding kinetic equations. However the stationary solution being only single one is not obvious at least because that the kinetic coefficients are bulky.

So it is not quite clear if the limitary distribution would be always the thermodynamic equilibrium. Actually an answer to this question was obtained only for a case of high temperature approximation (see [5], where this question has been discussed most openly), although the kinetic equations by themselves have been deduced for arbitrary temperature.

In the present work we will show that an assumption about short correlation times and large heat capacity is so strong that enables one to derive the kinetic equations without using the perturbation theory for spin-lattice interaction operator. The perturbation theory is applied only on a stage of calculating the kinetic coefficients.

The kinetic equations for the density matrix elements do not play the main role and direct derivation of the generalized Bloch equations for the expectation values of any set of physical quantities is possible answering only a weak condition of mutual orthogonality and completeness as well. The structure of kinetic coefficients is both general and "invariant". The kinetic equations for the density matrix elements appear to be only particular case of the general equations presented in a spatial operator basis.

The proof of the thermodynamic equilibrium established in spin subsystem has been performed in a simple general way and appears to be a mathematical consequence of existing a unity operator in the algebra of spin variables and nondegenerasy of the matrix of kinetic coefficients.

## 2. Derivation of the kinetic equations: approximation of short correlation times.

Let the small Latin letters $\hat{a}, \hat{b}, \hat{c}...$ indicate the spin operators. A set of all spin operators $R(S)$ forms an operator algebra (see, for example [11]), in other words, it has properties of a vector space, where a vector multiplying operation is determined. The operation of scalar production in $R(S)$ is defined by the relation:

$$\langle \hat{a} | \hat{b} \rangle = Tr_s \left( \hat{a}^* \hat{b} \right), \tag{1}$$

where $|\hat{b}\rangle$ denotes the spin operator $\hat{b}$ considered as a vector of $R(S)$, $\langle \hat{a}|$ represents the hermitian conjugate operator to $\hat{a}$, $Tr_s(...)$ denotes the trace operation over spin variables. After the operation $\langle \hat{a} | \hat{b} \rangle$ has been introduced, $R(S)$ becomes the normalized operator algebra and in particular case it posses all features of the Hermitian space. In physics this circumstance is indicated by term "Liouville space" or Liouville formalism.

Let choose an orthogonal basis $\hat{a}_0, \hat{a}_1, ...$ in $R(S)$ space. Then every spin operator $\hat{b}$ can be expressed as:

$$\hat{b} = \sum_n x_n \hat{a}_n, \tag{2a}$$

where

$$x_n = \frac{\langle \hat{a}_n | \hat{b} \rangle}{\langle \hat{a}_n | \hat{a}_n \rangle} = \frac{Tr_s \left( \hat{a}_n^* \hat{b} \right)}{Tr_s \left( \hat{a}_n^* \hat{a}_n \right)} \tag{2b}$$

and

$$\langle \hat{a}_n | \hat{a}_k \rangle = \delta_{nk} \langle \hat{a}_n | \hat{a}_n \rangle. \tag{2c}$$

The numbers $x_n$ are the coordinates of operator $\hat{b}$ in the basis $\hat{a}_0, \hat{a}_1, ...$ Further we consider that the basis vector $|\hat{a}_0\rangle \equiv \hat{a}_0$ is equal to a unity operator, i.e.

$$\hat{a}_0 = I. \tag{3}$$

From the orthogonality condition (2c) it follows that the operators $\hat{a}_1, \hat{a}_2, ...$ have zero trace:

$$\langle I | \hat{a}_n \rangle = Tr_s \left( \hat{a}_n \right) = 0. \tag{4}$$

The Hamiltonian of the whole system "spin + lattice" can be presented in a standard way:

$$\hat{H} = \hat{H}_0 + \hat{V} = \hat{H}_s + \hat{H}_L + \hat{V}, \tag{5}$$

where $\hat{H}_s$ is the spin Hamiltonian describing a spectrum of magnetic resonance;

$\hat{H}_L$ is the Hamiltonian of the lattice degrees of freedom,

$\hat{H}_0 = \hat{H}_s + \hat{H}_L$,

$\hat{V}$ is the Hamiltonian of the spin-lattice interaction.

The density matrix (or the statistical operator) of all system satisfy to the von Neumann equation:

$$i\hbar \frac{\partial}{\partial t} \hat{\rho}(t) = \left[ \hat{H}; \hat{\rho}(t) \right]. \tag{6}$$

In order to simplify the subsequent calculations we rewrite the von Neumann equation (6) in terms of the Liouville superoperator:

$$i\frac{\partial}{\partial t}\hat{\rho}(t) = \hat{L}_H \hat{\rho}(t), \qquad (7)$$

where Liouville operator can be defined by a relation:

$$\hat{L}_H \hat{\rho} \equiv \frac{1}{\hbar}\left[\hat{H};\hat{\rho}\right]. \qquad (7a)$$

The Liouville operator is the linear one acting in the algebra of all "usual" operators of the "spin + lattice" system. Sometimes such operators are called as a "superoperator" in order to stress this fact.

The total Hamiltonian $\hat{H}$, given by the expression (5), is time independent, therefore a formal solution of eq. (7) is:

$$\hat{\rho}(t) = \hat{S}(t)\hat{\rho}_0, \qquad (8)$$

with $\hat{S}(t) = \exp\{-i\hat{L}_H t\}$ -an evolution superoperator of all system and $\hat{\rho}_0$ is an initial value of the density matrix of all system.

Note, that we are treating dynamics both lattice and spin systems quantum mechanically, the lattice variables are operators and does not depend from time in the Shcrodinger representation, the spin Hamiltonian is assumed being static, i.e. no actions of external time dependent magnetic fields.

The evolution superoperator satisfies to the following equation:

$$\frac{d}{dt}\hat{S}(t) = -i\hat{L}_H \hat{S}(t), \qquad (9)$$

with the initial condition given by $\hat{S}(0) = I$.

Let us to consider the n-th basal spin operator $\hat{a}_n$. Knowing the state of system and the density matrix $\hat{\rho}(t)$ one can calculate the expectation value of physical quantity $\hat{a}_n$ at a moment of time $t$:

$$\langle \hat{a}_n(t) \rangle = Tr(\hat{a}_n \hat{\rho}(t)), \qquad (10)$$

where the trace operation is preformed over the all degrees of freedom, i.e. over spin and lattice variables.

Then we perform differentiation of both parts of the equation:

$$\frac{d}{dt}\langle \hat{a}_n(t) \rangle = Tr\left(\hat{a}_n \frac{1}{i}\hat{L}_H \hat{\rho}(t)\right). \qquad (11)$$

By replacing the operator $\hat{L}_H$ on the conjugate one in eq. (11) we can transform it to:

$$\frac{d}{dt}\langle \hat{a}_n(t)\rangle = Tr\left(\left(i\hat{L}_H \hat{a}_n\right)\hat{\rho}(t)\right). \tag{12}$$

Let us consider an interval of the time $\Delta t$. For the superoperator of evolution $\hat{S}(t+\Delta t)$ the relation

$$\hat{S}(t+\Delta t) = \hat{S}(\Delta t)\hat{S}(t) \tag{13}$$

is obvious.

Further, in accordance with the expansion of the basic Hamiltonian given by eq. (5), the Liouville superoperator $\hat{L}_H$ can be expressed in similar way:

$$\hat{L}_H = \hat{L}_0 + \hat{L}_V = \hat{L}_s + \hat{L}_L + \hat{L}_V, \tag{14}$$

where $\hat{L}_s, \hat{L}_L, \hat{L}_V$ are the contributions induced by the Hamiltonians $\hat{H}_s, \hat{H}_L, \hat{V}$ correspondingly.

After substituting the expression (14) into eq. (12) and using the fact as $\hat{a}_n$ to be merely spin operator, so that $\hat{L}_L \hat{a}_n = 0$, we can transforms it in a form:

$$\frac{d}{dt}\langle \hat{a}_n(t)\rangle = Tr\left(\left(i\hat{L}_s \hat{a}_n\right)\hat{\rho}(t)\right) + Tr\left(\left(i\hat{L}_V \hat{a}_n\right)\hat{S}(t)\hat{\rho}_0\right). \tag{15}$$

The superoperator $\hat{L}_s$ acts only on spin variables, so that the operator $\hat{L}_s \hat{a}_n$ belongs to the spin subspace of the Liouville space of the "spin + lattice" system. In accordance with the equations (2a) and (2b) we can expand into a series:

$$\hat{L}_s \hat{a}_n = \sum_k i\omega_{nk} \hat{a}_k, \tag{16}$$

the numbers

$$\omega_{nk} = \frac{Tr_s\left(\hat{a}_k^* \hat{L}_s \hat{a}_n\right)}{Tr_s\left(\hat{a}_k^* \hat{a}_k\right)} = \frac{1}{\hbar^2}\frac{Tr_s\left(\hat{a}_k^*\left[\hat{H}_s;\hat{a}_n\right]\right)}{Tr_s\left(\hat{a}_k^* \hat{a}_k\right)} \tag{16a}$$

form the frequency matrix.

Using eq. (16) we can transform eq. (15) to:

$$\frac{d}{dt}\langle \hat{a}_n(t)\rangle = \sum_k i\omega_{nk} \langle \hat{a}_k(t)\rangle + Tr\left(\left(i\hat{L}_V \hat{a}_n\right)\hat{S}(t)\hat{\rho}_0\right). \tag{17}$$

To get the kinetic equations from relation (16) we have to give them closed form, i.e. to express the second relaxation term in them as some function of observable expectation values $\langle \hat{a}_n(t)\rangle$. With this aid let us rewrite the expression (17) for the moment of time $t + \Delta t$:

$$\frac{d}{dt}\langle \hat{a}_n(t+\Delta t)\rangle = \sum_k i\omega_{nk} \langle \hat{a}_k(t+\Delta t)\rangle + Tr\left(\left(i\hat{L}_V \hat{a}_n\right)\hat{S}(t+\Delta t)\hat{\rho}_0\right). \tag{18}$$

It will be convenient to assign a special symbol to the second term in equation (18):

$$A_n(t+\Delta t) = Tr\left(\left(i\hat{L}_V \hat{a}_n\right)\hat{S}(t+\Delta t)\hat{\rho}_0\right) = Tr\left(\left(i\hat{L}_V \hat{a}_n\right)\hat{S}(\Delta t)\hat{\rho}(t)\right). \tag{19}$$

Let consider an equality:

$$\hat{S}(\Delta t)\hat{\rho}(t) \equiv \hat{S}\left((1-\alpha)\Delta t\right)\hat{S}\left(\alpha \Delta t\right)\hat{\rho}(t), \tag{20}$$

where $\alpha \ll 1$ a positive number.

Further we will consider situations, when the heat capacity of the lattice is larger than that of spin subsystem. If the lattice characteristic correlation time $\tau_0$ is shorter, than the characteristic spin-relaxation time $\tau_s$, we can find $\Delta t$ and $\alpha$ satisfying to

$$\tau_s \gg \Delta t \gg \alpha \Delta t \gg \tau_0. \tag{21}$$

During the time interval $\alpha \Delta t$ in accordance with (21) both spin subsystem and lattice evolve actually independently. The Gibbs equilibrium distribution is establishing relative to the lattice degrees of freedom. Since the heart capacity of the lattice is much larger than the spin subsystem we can neglect by a time dependence of lattice temperature. Therefore we can transform the expression (20) to the following:

$$\hat{S}(\Delta t)\hat{\rho}(t) \equiv \hat{S}\left((1-\alpha)\Delta t\right)\hat{\rho}_s(t+\alpha \Delta t)\hat{\rho}_L^{eq}, \tag{22}$$

where $\hat{\rho}_s(t)$ is purely spin density matrix,

$\hat{\rho}_L^{eq}$ is the Gibbs equilibrium density matrix of lattice. Taking into account, that $\alpha \ll 1$, we have:

$$\hat{\rho}(t+\Delta t) \equiv \hat{S}(\Delta t)\hat{\rho}_s(t)\hat{\rho}_L^{eq} \tag{23}$$

with accuracy of order $\alpha$.

Replacing in eq. (19) the superoperator $\hat{S}(\Delta t)$ on the conjugate one leads to:

$$A_n(t+\Delta t) = Tr\left(\hat{\rho}_s(t)\hat{\rho}_L^{eq}\left(\hat{S}^*(\Delta t)i\hat{L}_V \hat{a}_n\right)\right). \tag{24}$$

Then we define a new operator that is purely spin:

$$\hat{A}_n(\Delta t) = Tr_L\left(\hat{\rho}_L^{eq}\left(\hat{S}^*(\Delta t)i\hat{L}_V \hat{a}_n\right)\right), \tag{25}$$

where $Tr_L$ denotes the trace operation over the lattice variables. An obvious relation takes place

$$A_n(t+\Delta t) = Tr_s\left(\hat{\rho}_s(t)\hat{A}_n(\Delta t)\right). \tag{26}$$

Note that operators $\hat{a}_k$ form a basis in algebra of spin operators. The same behaviors they display after transformation to the interaction representation, the Dirac representation, at the moment of time $\Delta t$. We define these new operators $\hat{\tilde{a}}_k(t)$ by the following expression:

$$\hat{\tilde{a}}_k(\Delta t) = \hat{S}_0^*(\Delta t)\hat{a}_k \equiv \exp\left\{i\frac{\hat{H}_0 \Delta t}{\hbar}\right\} \hat{a}_k \exp\left\{-i\frac{\hat{H}_0 \Delta t}{\hbar}\right\} =$$
$$\exp\left\{i\frac{\hat{H}_s \Delta t}{\hbar}\right\} \hat{a}_k \exp\left\{-i\frac{\hat{H}_s \Delta t}{\hbar}\right\} \tag{27}$$

where superoperator $\hat{S}_0^*(\Delta t)$ defines the transfer into the interaction representation.

Note that the superoperator $\hat{S}_0(\Delta t)$ conjugated to $\hat{S}_0^*(\Delta t)$ describes the evolution superoperator induced by the Hamiltonian $\hat{H}_0 = \hat{H}_s + \hat{H}_L$.

The operator $\hat{A}_n(\Delta t)$ in accordance with the definition (25) is purely spin operator. Therefore we can represent it as the linear combination of basis of vectors $\hat{\tilde{a}}_0(t), \hat{\tilde{a}}_1(t),\ldots$:

$$\hat{A}_n(\Delta t) = \sum_k w_{nk}(\Delta t) \hat{\tilde{a}}_k(\Delta t), \tag{28}$$

where $w_{nk}(\Delta t) = \dfrac{Tr_s\left(\hat{\tilde{a}}_k^* \hat{A}_n(\Delta t)\right)}{Tr_s\left(\hat{a}_k^* \hat{a}_k\right)}$. \hfill (29)

After substitution relation (25) for the operator $\hat{A}_n(\Delta t)$ into eq. (29) and replacing the superoperator $\hat{S}_0^*(\Delta t)$ by the conjugate one and taking into account that the equilibrium density matrix of the lattice isn't change after action of the superoperator $\hat{S}_0^*(\Delta t)$, we can get:

$$w_{nk}(\Delta t) = \frac{Tr\left((i\hat{L}_V \hat{a}_n)(\hat{S}(\Delta t)\hat{S}_0^*(\Delta t)\hat{a}_k^* \hat{\rho}_L^{eq})\right)}{Tr_s\left(\hat{a}_k^* \hat{a}_k\right)}. \tag{30}$$

Note that in the nominator of eq. (30) *Tr* operation is performed over all the spin and lattice degrees of freedom of the system whereas in the dominator this operation concerns only spin variables.

Consider an equality:

$$\hat{S}(t_2 - t_1) = \hat{S}_0(t_2) \hat{\tilde{S}}_V(t_2; t_1) \hat{S}_0^*(t_1), \tag{31}$$

where the evolution superoperator $\hat{\tilde{S}}_V(t_2; t_1)$ describes the evolution of density matrix in the time interval from $t_1$ up to $t_2$ in the interaction representation.

The evolution superoperator $\hat{\tilde{S}}_V(t_2; t_1)$ is determined by Dyson chronological exponent:

$$\hat{\tilde{S}}_V(t_2;t_1) = \hat{T}\exp\left\{-i\int_{t_1}^{t_2}\hat{\tilde{L}}_V(\tau)d\tau\right\}, \tag{32}$$

where the Liouville superoperator is produced by the operator $\hat{\tilde{V}}(\tau) = \hat{S}_0^*(\tau)\hat{V}$ of the spin-lattice relaxation in the Dirac representation.

Let us rewrite equation (31) for the moments of time $t_2 = 0$ and $t_1 = -\Delta t$

$$\hat{S}(\Delta t) = \hat{\tilde{S}}_V(0;-\Delta t)\hat{S}_0^*(\Delta t), \tag{33}$$

where we are using the equality $\hat{S}_0^*(-\Delta t) = \hat{S}_0(\Delta t)$.

Substituting the equality (33) in eq. (30) we get:

$$w_{nk}(\Delta t) = \frac{Tr\left[\left(i\hat{L}_V\hat{a}_n\right)\left(\hat{\tilde{S}}_V(0;-\Delta t)\hat{a}_k^*\hat{\rho}_L^{eq}\right)\right]}{Tr_s\left(\hat{a}_k^*\hat{a}_k\right)}. \tag{34}$$

Then using symmetry of the equation (7) relatively a time translation operation, we can transform eq. (34) to

$$w_{nk}(\Delta t) = \frac{Tr\left[\left(i\hat{L}_V\hat{a}_n\right)\left(\hat{\tilde{S}}_V(\Delta t;0)\hat{a}_k^*\hat{\rho}_L^{eq}\right)\right]}{Tr_s\left(\hat{a}_k^*\hat{a}_k\right)}. \tag{35}$$

The superoperator $\hat{\tilde{S}}_V(\Delta t;0)$ can be expressed as

$$\hat{\tilde{S}}_V(\Delta t;0) = I + \int_0^{\Delta t} d\tau \frac{d}{d\tau}\hat{\tilde{S}}_V(\tau;0) = I + \int_0^{\Delta t} d\tau \frac{1}{i}\hat{\tilde{L}}_V(\tau)\hat{\tilde{S}}_V(\tau;0). \tag{36}$$

The spin-lattice interaction operator always can be chosen in such a way that

$$Tr_L(\hat{V}\hat{\rho}_L^{eq}) = 0. \tag{37}$$

Indeed if $Tr_L(\hat{V}\hat{\rho}_L^{eq}) \neq 0$, it would be by construction a purely spin dependent operator and can be included in the spin Hamiltonian $\hat{H}_s$ and instead the Hamiltonian $\hat{V}$ one can use everywhere it's fluctuating part $\delta\hat{V} = \hat{V} - Tr_L\left(\hat{V}\hat{\rho}_L^{eq}\right)$.

Substituting eq. (36) into (35) and using eq. (37) yields for $w_{nk}(\Delta t)$ after simple transformations:

$$w_{nk}(\Delta t) = \frac{\int_0^{\Delta t} Tr\left((\hat{L}_V\hat{a}_n)(\hat{\tilde{L}}_V(\tau)\hat{\tilde{S}}_V(\tau;0)\hat{a}_k^*\hat{\rho}_L^{ea})\right)d\tau}{Spur_s\left(\hat{a}_k^*\hat{a}_k\right)}. \tag{38}$$

The superoperator $\hat{\tilde{S}}_V(\tau;0)$ from eq. (38) contains the oscillating spin variables with the characteristic frequency $\omega_0$ and the lattice functions depending on time. After $Tr$ operation being performed, the equilibrium lattice correlation functions arise and they are changing during a characteristic correlation time $\tau_0$. Suppose that the spin system obeys to the condition of short correlation times

$$\tau_S \gg \min\{\omega_0^{-1}, \tau_0\}, \tag{39}$$

where $\tau_S$ is a typical time of spin relaxation. Then the time interval $\Delta t$ can be chosen to follow a condition

$$\tau_S \gg \Delta t \gg \min\{\omega_0^{-1}; \tau_0\}. \tag{40}$$

If one expands the superoperator $\hat{\tilde{S}}_V(\tau;0)$ from (38) in Tailor's series by $\hat{\tilde{L}}_V(\tau)$ and perform integration over time, he finds the terms practically independent on $\Delta t$ and the terms proportional to $(\Delta t)^n$, where $n \geq 1$, as well. Let us to call a sum of the former terms as a regular part of eq. (38) and a sum of the latter as a singular part of this expression. The regular part of eq. (38) is connected with decaying lattice correlation functions appearing after performing $Tr$ operation. The regular part can be estimated to be of $\tau_S^{-1}$ over the order of magnitude.

The singular part corresponds to the non decaying correlation functions, which containing in eq. (38) in terms of order $\hat{V}^4$ and higher. In typical situations the reasonable estimation of the singular part is of order of magnitude $\Delta t \tau_S^{-2}$. If it is true, we can neglect by singular part of eq. (38) in comparison with its regular part taking into account relation (40). So eq. (38) can be rewritten

$$w_{nk} = \frac{\int_0^\infty Tr\left((\hat{L}_V \hat{a}_n)(\hat{\tilde{L}}_V(\tau)\hat{\tilde{S}}_V(\tau)\hat{a}_k^* \hat{\rho}_L^{ea})\right)_{\text{Re}\,g} d\tau}{Tr_s\left(\hat{a}_k^* \hat{a}_k\right)}, \tag{41}$$

where the index $\text{Re}\,g$ denotes, that only the regular part of the integral must be under consideration.

Note, that appearance of divergent (or singular) terms in contributions of perturbation theory in terms higher than second order is an usual situation for all systems with infinitely (macroscopically) large degrees of freedom. A regularization procedure any time is necessary in that cases, the quantum field theory is most known example. In our case the divergence, if one would formally treat the expression (41), appear due to upper limit of integration. Actually in the expression (41) instead infinity in upper limit of integration should be $\varepsilon w_{nk}^{-1}$ with a $\varepsilon \ll 1$. In this case no divergent terms in the expression (41), which can consider as a transcendental equation for the kinetic coefficient $w_{nk}$. If there exist it's solution for $\varepsilon \ll 1$, which dependence from $\varepsilon$ is negligible, then the approximation of short correlation times is correct.

In this connection we have to mention the papers by Aminov [13,14], in which the contributions from the terms of order $\hat{V}^4$ to the kinetic coefficients were given a detailed scrutiny for the case of the spin-lattice relaxation in solids. Some results of this works appeared to be proof that the contribution of the singular part must be excluded.

In general case, strictly speaking, a definition of the short correlation time approach includes the existence of the time interval $\Delta t$, that answers to the condition (40), and negligible contribution of the singular part in comparison with that of the regular part, as well. So our assumption preceding eq. (41) must be considered as to be purely heuristic character that makes our definition of the short correlation time approach reasonable. If discussed assumptions are not satisfied, the kinetic equations describing evolution of spin variables can not be properly approximated by ordinary differential equations of first order and would have essentially integro-differential character. The memory function formalism, based on projection operator technique is more adequate.

Returning to derivation of kinetic equations we transfer formula (18) by means of eq. (19), (25), (35) to

$$\frac{d}{dt}\langle \hat{a}_n(t+\Delta t)\rangle = \sum_k i\omega_{nk} \langle \hat{a}_k(t+\Delta t)\rangle + \sum_k w_{nk} Tr_s\left(\hat{S}_0(\Delta t)\hat{a}_k \hat{\rho}_s(t)\right). \tag{42}$$

Then since inequality (40) a following relation is valid

$$\hat{S}_0(\Delta t)\hat{\rho}_S(t) \cong \hat{S}(\Delta t)\hat{\rho}_S(t) = \hat{\rho}_S(t+\Delta t). \tag{43}$$

Note that the approximate equality is true within the order of magnitude $\tau_0/\tau_s \ll 1$.

At last we can replace the operator $\hat{S}_0^*(\Delta t)$ in eq. (42) by the conjugate one and then after using approximation (43) and redenoting the value $t+\Delta t$ as t we find that

$$\frac{d}{dt}<\hat{a}_n(t)> = \sum_k i\omega_{nk} <\hat{a}_k(t)> + \sum_k w_{nk} <\hat{a}_k(t)>. \tag{44}$$

In fact we already derived the kinetic equations for the expectation value of basal operators $<\hat{a}_n(t)>$. Let us show that the equilibrium values $<\hat{a}_n(t)> = <\hat{a}_n>_{eq}$ the stationary solutions of eq. (44). Firs of all note that the equilibrium density matrix of the hole system does not change itself under the influence of the evolution superoperator $\hat{S}(t)$

$$\hat{S}(t)\hat{\rho}_{eq} = \hat{\rho}_{eq} \cong \hat{\rho}_S^{eq}\hat{\rho}_L^{eq}. \tag{45}$$

where $\hat{\rho}_S^{eq} = \dfrac{1}{Z_S}\exp\{-\beta\hat{H}_S\}$ is the equilibrium density matrix of spin subsystem.

Then acting on both sides of the eq. (45) by the superoperator $\hat{S}_0^*(t)$ we have

$$\hat{S}_0^*(t)\hat{S}(t)\hat{\rho}_{eq} = \hat{\tilde{S}}_V(t)\hat{\rho}_{eq} \simeq \hat{\tilde{S}}_V(t)\hat{\rho}_S^{eq}\hat{\rho}_L^{eq} \simeq \hat{\rho}_S^{eq}\hat{\rho}_L^{eq}. \tag{45a}$$

The set of operators $\hat{a}_0^*, \hat{a}_1^*, \hat{a}_2^*, \ldots$ forms a basis in the space $R(S)$ as the set $\hat{a}_0, \hat{a}_1, \hat{a}_2, \ldots$ does. The spin equilibrium density matrix can be represented as a linear combination

$$\hat{\rho}_S^{eq} = \sum_k \frac{<\hat{a}_k>_{eq}}{Tr_s(\hat{a}_k^*\hat{a}_k)}\hat{a}_k^*. \tag{46}$$

Differentiating both parts of eq. (45a) over time yields the result

$$\hat{\tilde{L}}_V(t)\hat{\tilde{S}}_V(t)\hat{\rho}_S^{eq}\hat{\rho}_L^{eq} \simeq 0. \tag{47}$$

Note, that the eq. (47) gives correct results for the expectation values of spin variables, with accuracy of order $\beta\|\hat{V}\|$, where $\beta = 1/k_B T$ is inverse temperature, $\|\hat{V}\|$ is norm of the spin-lattice Hamiltonian. This can bee found using Kubo-Tomita decomposition for the operator $\exp\{\beta(\hat{H}_s + \hat{H}_L)\}\exp\{-\beta(\hat{H}_s + \hat{H}_L + \hat{V})\}$ containing in a hidden way in right sids of the eqs. (45) and (45a). Therefore our derivation at this stages needs rather week assumption about high temperature approximation relatively the spin-lattice relaxation Hamiltonian, i.e. $\beta\|V\| \ll 1$.

In principle, mentioned corrections can be systematically taken into account by introducing effective temperature dependent and time independent spin, lattice and spin-lattice interaction Hamiltonians defined by the following relations correspondingly:

$$\begin{aligned}\hat{H}_s^* &\equiv -k_B T \ln\left(Tr_Q(\hat{\rho}^{eq})\right) \\ \hat{H}_L^* &\equiv -k_B T \ln\left(Tr_s(\hat{\rho}^{eq})\right) \\ \hat{H}_s + \hat{H}_L + \hat{V} &\equiv \hat{H}_s^* + \hat{H}_L^* + \hat{V}^*\end{aligned} \tag{47a}$$

Actually discussed corrections have mainly an academic interest, because for any known situation $\beta\|\hat{V}\| \ll 1$ even at temperatures, when the approximation of short correlation times starts to be incorrect due to a slow motions of the lattice, for example for transverse relaxation. For the longitudinal relaxation, i.e. the spin-lattice relaxation, the BRW equations applicable even at small temperatures, if the spin-lattice interaction Hamiltonian is smaller than the spin

Hamiltonian, i.e. $\|\hat{V}\| \ll \|\hat{H}_s\|$ and heat capacity of the lattice is much larger, than heat capacity of the spin-system. In this situations we can consider a commutative subalgebra created only by operators of z-projections of spin vectors. All another spin degrees of freedom can be included in an extended lattice. A role of short correlation time in this case is playing inverse resonance frequency $\omega_0^{-1}$.

Note also, that if the operator of the spin-lattice relaxation commute with the Hamiltonian of all system, i.e. $\left[\hat{V}; \hat{H}\right] = \left[\hat{V}; \hat{H}_L + \hat{H}_s\right] = 0$, then the expression (47) is exact, therefore all mentioned corrections connected with nonsecular part of the Hamiltonian $\hat{V}$. It is is oscillating and fast time dependent in at the interaction representation, what is a general reason, why mentioned corrections can not be large when an approximation of short correlation times is correct.

Now one should remember that operator $\hat{a}_0$ is equal to unity operator . i.e. $\hat{a}_0 = I$. Inserting the expansion (46) into relation (47) and gathering all terms with $k \geq 1$ in the right-hand part one find that

$$\frac{\hat{\hat{L}}_V(t)\hat{\hat{S}}_V(t)\hat{\rho}_L^{eq}}{Spur_s(\hat{a}_0^*\hat{a}_0)} = -\sum_{k \geq 1} \frac{<\hat{a}_k>_{eq}}{Spur_s(\hat{a}_k^*\hat{a}_k)} \hat{\hat{L}}_V(t)\hat{\hat{S}}_V(t)\hat{a}_k^*\hat{\rho}_L^{eq} . \qquad (48)$$

Then we single out the terms with $\hat{a}_0$ in each sum in eq. (44) and by means of the relation (48) transform them to

$$\frac{d}{dt}<\hat{a}_n(t)> = \sum_k i\omega_{nk} <\hat{a}_k(t)> + \sum_{k \geq 1} w_{nk}(<\hat{a}_k(t)> - <\hat{a}_k>_{eq}) . \qquad (49)$$

It is obvious that the equilibrium values $<\hat{a}_n(t)> = <\hat{a}_n>_{eq}$ belong to a kernel of the frequency matrix $\omega_{nk}$, i.e. $\sum_k \omega_{nk} <\hat{a}_k>_{eq} = 0$. For this purpose it is necessary to use relations (16a), (46) and commutativity of the spin Hamiltonian $\hat{H}_S$ with $\hat{\rho}_S^{eq}$. Note also, that $\omega_{0k} \equiv 0$, because $\hat{a}_0 = I$, see eq. (16a) .

Thus we can transform the eq. (49) to

$$\frac{d}{dt}<\hat{a}_n(t)> = \sum_k (i\omega_{nk} + w_{nk})(<\hat{a}_k(t)> - <\hat{a}_k>_{eq}) . \qquad (50)$$

For $n = 0$ the first of equations (50) reduces to the identity $0 \equiv 0$. For all other $n$ these equations describe free relaxation of expectation values of spin operators $<\hat{a}_n(t)>$. It is evident that the equilibrium values $<a_n(t)> = <a_n>_{eq}$ form the stationary solution of kinetic equations (50). Uniqueness of solutions reduces to a problem about undegeneracy of the whole matrix of kinetic coefficients $i\omega_{nk} + w_{nk}$, where $n \geq 1$. The problem about existing the limiting values $\lim_{t \to \infty} <\hat{a}_n(t)>$ is closely connected with sign of real part of the eigenvalues of the matrix $i\omega_{nk} + w_{nk}$. Let us denote the mentioned eigenvalues as $\lambda_n$. So if the spin-lattice relaxation operator $\hat{V}$ is such that $\text{Re}(\lambda_n) < 0$ for all $n \geq 1$, the limiting distribution exist for all initial conditions and it is the Gibbs equilibrium distribution. If the matrix of kinetic coefficients is degenerate, this means that spin-lattice relaxation Hamiltonian $\hat{V}$ does not properly mix spin and lattice degrees of freedom, therefore the spin subsystem starting from arbitrary initial state should not evolve in course dynamical evolution to equilibrium state with temperature of the lattice.

When $n \neq k$, in the case of multiparticle spin-lattice relaxation the kinetic coefficients $w_{nk}$ describe the cross-relaxation transitions of spin-diffusion type, of course, if $\hat{a}_n$ and $\hat{a}_k$ are the operators relating to different spins. In general case the kinetic coefficients $w_{nk}$ defined by relations (41) are complex. Therefore the imaginary part $\text{Im}\,\lambda_n$ eigenvalues of the matrix $i\omega_{nk} + w_{nk}$ differs from the eigenvalues of the frequency matrix $i\omega_{nk}$. This difference is due to dynamical shift of the resonance frequencies influenced by spin-lattice relaxation.

Subsequent calculations of kinetic coefficients $w_{nk}$ can be performed by applying the perturbation theory for computation the evolution of superoperator

$$\hat{\tilde{S}}_V(t) = \hat{T} \exp\left\{-i\int_0^t \hat{\tilde{L}}_V(t)dt\right\} = 1 - i\int_0^t \hat{\tilde{L}}_V(t_1)dt_1 - \int_0^t dt_2 \hat{\tilde{L}}_V(t_2)\int_0^{t_2} dt_1 \hat{\tilde{L}}_V(t_1) + \ldots \quad (51)$$

In zero approximation: $\hat{\tilde{S}}_V(t) = I$. Using the definition of $\hat{\tilde{L}}_V(t)$ and the relation (7a) we obtain from eq. (41)

$$w_{nk} = \frac{1}{\hbar^2} \frac{\int_0^\infty dt\, Tr\left([\hat{V};\hat{a}_n][\hat{\tilde{V}}(t);\hat{a}_k^* \hat{\rho}_L^{ea}]\right)}{Tr_s\left(\hat{a}_k^* \hat{a}_k\right)}. \quad (52)$$

This result takes into account the contributions to the kinetic coefficients $w_{nk}$ arising in the perturbation theory of the second order for the spin-lattice operator $\hat{V}$. In this case the expression (52) does not contains the singular part mentioned above. In the case of spin-lattice relaxation in solids it is just the same as one takes into account both one phonon and multi-phonon processes of first order.

Considerations of next terms of the expansion (51) that give non-zero contribution to $w_{nk}$ is equivalent to the study of the multi-phonon processes of higher order [13,14] . The singular terms raised in this case in eq. (38) due to multi-phonon processes connected with the "resonant" phonons. Aminov showed [13,14] by a direct calculations that in the approximation of short correlation times the singular terms must be excluded from kinetic coefficients. In brief the physical essence is that the regular terms are connected with virtual multi-phonon processes, but the singular terms are connected with real multi-phonon processes. So the singular terms are reducible and they are taken into account in the structure of kinetic equations by the regular terms of lower order over the spin-lattice interaction $\hat{V}$.

In the case of NMR relaxation in liquids expression (52) is a compact mathematical reformulation of the Bloch-Wangsness-Redfield theory.

## 3. Simplest illustration.

Let us consider relaxation of a spin $I = \frac{1}{2}$ in an effective molecular magnetic field. The spin operators $\hat{a}_0 = I, \hat{a}_1 = \hat{I}_x, \hat{a}_2 = \hat{I}_y, \hat{a}_3 = \hat{I}_z$ can be chosen as basal operators. The Hamiltonian of the spin-lattice interaction of a spin with an effective molecular magnetic field is

$$\hat{V} = -\gamma \hbar \vec{H}^*(\{\vec{r}_i\}) \cdot \vec{I}, \tag{53}$$

where $\vec{H}^*(\{\vec{r}_i\})$ is the effective molecular field, depending on spacing between the particles of lattice, $\gamma$ is the gyromagnetic ratio.

We can express the spin Hamiltonian in a way

$$\hat{H}_S = \hbar \omega_0 \hat{I}^z, \tag{54}$$

where $\omega_0$ is the resonance frequency.

After performing elementary $Tr$ operations we obtain according to eq. (52)

$$w_{nk} = \gamma^2 \int_0^\infty dt \{ \langle \tilde{H}_k^*(t) \tilde{H}_n(0) \rangle_{eq} - \delta_{nk} \langle \vec{\tilde{H}}^*(t) \vec{\tilde{H}}(0) \rangle_{eq} \}, \tag{55}$$

where

$$\left. \begin{array}{l} \tilde{H}_x^*(t) = Cos\omega_0 t H_x^*(t) + Sin\omega_0 t H_y^*(t) \\ \tilde{H}_y^*(t) = -Sin\omega_0 t H_x^*(t) + Cos\omega_0 t H_y^*(t) \\ \tilde{H}_z^*(t) = H_z^*(t) \end{array} \right\}, \tag{56}$$

and the brackets $<...>_{eq}$ denote the averaging over the equilibrium density matrix of lattice $\hat{\rho}_L^{eq}$. Note that the time dependence of equilibrium correlation functions that are the components of the effective magnetic field $\vec{H}^*(\{\vec{r}_i(t)\}) \equiv \vec{H}^*(t)$ is caused by transition to the interaction representation with the Hamiltonian $\hat{H}_0 = \hat{H}_s + \hat{H}_L$.

The terms including $\cos(\omega_0 t)$ correspond to the relaxation processes and the terms with $\sin(\omega_0 t)$ describe the dynamical shift of the resonance frequency. In the simplest isotropic case, when

$$< H_\alpha^*(t) H_\beta^*(0) > = \frac{1}{3} \delta_{\alpha\beta} < \vec{H}^*(t)\vec{H}^*(0) >_{eq}, \tag{57}$$

the standard results are obtained: kinetic of spin relaxation is characterized by two relaxation times $T_1$ and $T_2$:

$$\frac{1}{T_1} = -w_{33} = \frac{2}{3}\gamma^2 \int_0^\infty dt \cos(\omega_0 t) \langle \vec{H}^*(t)\vec{H}^*(0) \rangle_{eq}, \tag{58}$$

$$\frac{1}{T_2} = -w_{11} = -w_{22} = \frac{1}{3}\gamma^2 \int_0^\infty dt \langle \vec{H}^*(t)\vec{H}^*(0) \rangle_{eq} + \frac{1}{2}\frac{1}{T_1}. \tag{59}$$

If the time dependent correlation function $\langle \vec{H}^*(t)\vec{H}^*(0) \rangle_{eq}$ is monotonically decaying and for any resonance frequencies

$$\frac{1}{T_2} \geq \frac{1}{T_1}. \tag{60}$$

The dynamical shift of resonance frequency is

$$\Delta\omega_0 = -w_{12} = -\frac{1}{3}\gamma^2 \int_0^\infty dt \sin(\omega_0 t) \langle \vec{H}^*(t)\vec{H}^*(0) \rangle_{eq}. \tag{61}$$

The equations (50) coincide with classical Bloch equations [5-8]. If we are deal with enough complex system, for examples, liquids in porous system, membranes, liquid crystals and so on, the correlation matrix $< H_\alpha^*(t) H_\beta^*(0) >$ can appear to be nonscalar. The kinetic equations (50) describe a time evolution of three different spin components. In general case here At low resonance frequencies, in a general case, here can be three independent relaxation modes and become possible situations, when inequality (60) is incorrect, i.e. $\frac{1}{T_2} < \frac{1}{T_1}$.

Now let us consider a proof that all known relations of the BWR theory appear to be a particular case of eq. (50) when the approximation (52) is used for kinetic coefficients $w_{nk}$. We denote the whole set of quantum numbers of spin subsystem as $\nu$. It is obvious that the spin operators $\hat{P}_{\nu\nu'} \equiv |\nu\rangle\langle\nu'|$, forms a basis in the space of spin operators. So we can consider operators $\hat{P}_{\nu\nu'}$ as the operators $\hat{a}_n$, holding that $n$ indicates double index $\nu\nu'$.

The kinetic equations (50) appear to be the equations for the evolution of matrix elements of spin density matrix in the laboratory frame. Then, if one presents the operator $\hat{V}$ as a sum of products of spin and lattice operators, the commutations in relation (52) can be performed in evident way and the trace operation can be expressed as a sum of corresponding matrix elements. Comparison of bulky and noninvariant expressions with results of the BWR theory convinced that they are identical.

In recent paper [15] it was shown that Redfield [4] and Torrey [16] approaches for treating of magnetic-field induced spin relaxation are identical. This conclusion automatically follow from our general derivation of the kinetic equations (52), which did not use any specific forms for a propagator describing spatial displacements of the spin. An experimental importance of this mechanism relaxation was recently demonstrated in paper [17].

## 4. Conclussion.

Both equations (50) and (41) give a general and complete solution of the problem of constructing the kinetic equations and calculating the kinetic coefficients for spin variables in the approximation of short correlation times and large heat capacity of the lattice. The expansion in the Liouville superoperator $\hat{\tilde{L}}_V(t)$ series (51) caused by the Hamiltonian spin-lattice interaction in the Dirac representation allow to calculate the kinetic coefficients $w_{nk}$ with necessary accuracy. In particular, obtained kinetic equations can be applied for describing relaxation of multi-spin system relaxation. The eq. (50) involves all contributions of the second order over the spin-lattice interaction Hamiltonian $\hat{V}$ and represents a new invariant and compact mathematical reformulating the BWR theory convenient for performing specific calculations (compare, for example, discussions of BWR theory in papers [12, 18-21] and it's interesting applications).

### Acknowledgments.
The author is grateful for useful discussions to L.K. Aminov and R. Kimmich.

**References.**

1. R.K.Wangsness, F. Bloch, Phys.Rev. **89**, 728 (1953).
2. F.Bloch F., Phys.Rev. **102,** 104 (1956).
3. F.Bloch , Phys.Rev. **105,** 1204 (1957).
4. A.J.Redfield , IBM Journal,,**1**, 19 (1957).
5. A.Abraham, *The principles of nuclear magnetism.* (Clarendon Press, Oxford 1961).
6. C.P.Slichter , *Principles of magnetic resonance.* (3$^{rd}$ edn. Springer-Verlag, Berlin, Heiderberg. New York 1992).
7. I.V. Alexandrov, *Theory of magnetic relaxation.* (Science ,Moscow, (in Russian) 1975).
8. S.A.Al'tshuler, B.M. Kosirev, *Electron paramagnetic resonance in compounds of transition elements.* (Halsted Press 1974).
9. K.M.Salikhov , A.G.Semenov, Yu.D.Tzvetkov, *Electron spin echo and its applications*, (Science, Novosibirsk 1978).
11. O .Bratteli, D.W. Robinson, *Operator algebras and quantum statistical mechanics*, (Springer-Verlag, Berlin, Heiderberg. New York 1974).
12. M. Goldman, J. Magn. Reson. **149**, 160 (2001) .
13. L.K.Aminov, Phys. Stat.Sol.(b), **50**, 405 (1972).
14. L.K. Aminov, Soviet Physics JETP, **40**, 41 (1974).
15. R.Golub, Ryan M.Pohm, and C.M.Swank, Phys.Rev. A**83** 023402-1 (2011).
16. H.C.Torrey, Phys. Rev. **104**, 563 (1953).
17. M.S. Tagirov, E.M. Alakshin, R.R. Gazizulin, A.V. Egorov, A.V. Klochkov · S.L., Korableva · V.V., Kuzmin · A.S., Nizamutdinov K., Kono · A., Nakao · A.T., Gubaidullin, J. Low Temp. Phys. **162**, 645 (2011).
18. Brueschweiler, R; Chem. Phys. Lett., **270**, 217 (1997).
19. Skrynnikov, N.R; Bruschweiler, R; Chem. Phys. Lett. 1997, **281**, 239 (1997).
20. Magusin, PCMM; Veeman, WS; J. Magn. Reson. **143**, 243 (2000).
21. Kuprov, I; Wagner-Rundell, N; Hore, PJ; J. Magn. Reson. **184**, 196 (2007).
22. Short version of this paper is published in: Fatkullin N (1994) Magnetic resonance and related phenomena. Extended abstracts of the XXVII-th Congress Ampere, Kazan, 235-236.